U.S. Core Inflation: A Wavelet Analysis

by

Kevin Dowd[*]

Nottingham University Business School

Phone: +44 115 846 6682; Fax: +44 115 846 6667

Email: Kevin.Dowd@Nottingham.ac.uk.

and

John Cotter

University College Dublin,

Phone +353 1 7168900

Email: john.cotter@ucd.ie.

---

[*] Kevin Dowd is Professor of Financial Risk Management at Nottingham University Business School. John Cotter is Director of the Centre for Financial Markets, Smurfit School of Business, University College Dublin. They thank Tim Cogley, Ramo Gençay, Julie K. Smith and two anonymous referees for helpful comments on an earlier draft. They also thank Julie Smith for permission to use her data.

Name and mailing address of corresponding author:

Kevin Dowd

Nottingham University Business School,

Jubilee Campus,

Wollaton Rd,

Nottingham NG8 1BB,

United Kingdom.

Abstract

This paper proposes the use of wavelet methods to estimate U.S. core inflation. It explains wavelet methods and suggests they are ideally suited to this task. Comparisons are made with traditional CPI-based and regression-based measures for their performance in following trend inflation and predicting future inflation. Results suggest that wavelet-based measures perform better, and sometimes much better, than the traditional approaches. These results suggest that wavelet methods are a promising avenue for future research on core inflation.

# 1. INTRODUCTION

Monetary economists have long understood that some measures of inflation are more important, or more revealing, than others, and the distinction between headline inflation and underlying, or core, inflation, has been recognised for many years. However, interest in core inflation took off only in the 1990s.[1] As inflation targeting became more widespread, central bankers became more concerned to find the 'right' inflation rate to target, or to use as a monetary policy indicator; and as inflation fell, disentangling inflation signals from inflation noise became more critical to successful monetary policy decision-making: to paraphrase Blinder (1997, p. 158), when inflation is high, the central bank knows it has to lower inflation, regardless of how it is measured; but when inflation is low, different measures of inflation can suggest different monetary policy decisions, and correctly extracting the inflation signals becomes critical.

The literature on core inflation suggests two alternative definitions of the term. Some authors define core inflation as an inflation measure that represents underlying or trend inflation (e.g., Bryan *et al*., 1997, Cecchetti, 1997), whilst others

---

[1] The term 'core inflation' goes back at least to Eckstein (1981, p. 7), who defined it as the "trend increase in the cost of the factors of production". However, modern usage of the term is somewhat different from Eckstein's original definition, and tends to focus more on trend inflation or inflation prediction. All discussions of core inflation, in turn, take place against the background of a

define core inflation as a leading indicator of future inflation (e.g., Blinder, 1997, Smith, 2004). In practice, we would often like a core inflation measure that is compatible with both definitions – namely, is a good representation of underlying trend inflation and is a good indicator of future inflation. Thus we analyse two performance criteria (and separate associated test methodologies) to determine whether the candidate core inflation measures are able to match the alternative definitions of core inflation (i.e., in terms of tracking a trend and predicting future inflation).[2]

We can also think of core inflation as follows. We start with a given original or 'parent' inflation series, whose 'core' it is that we are seeking. We can then regard core inflation as some series that is closely related to the 'parent' series, but that also satisfies certain desirable properties such as capturing the trend, predicting the 'parent' series, and so forth.

Much of the literature on core inflation in the U.S. has taken the 'parent' inflation rate to be the CPI and examined measures of core inflation derived directly

---

longstanding debate on inflation measurement that goes back to Irving Fisher and earlier. For more on these issues and the origin of the notion of core inflation, see, e.g., Wynne (1997) and Roger (1998).

[2] Ideally, we might also want a measure of core inflation to satisfy a number of additional criteria (see Wynne (1999) for a detailed list): to be robust to ancillary assumptions, credible, verifiable, transparent, have some economic-theoretical basis (e.g., in monetary theory, as in Quah and Vahey, 1995), remove seasonality (as in Cecchetti, 1995), be implementable in real-time (and produce estimates of core inflation that do not change as new inflation observations become available), and produce results that are easy to communicate (see Rogers, 1998). However, no measure of core inflation satisfies all these criteria, and one is inevitably forced to compromise and accept core inflation measures that only satisfy some of them. We return to some of these issues further below.

from the components that make up the aggregate CPI index.[3] The most widely used such measure is 'CPI less food and energy' – inflation measured on the basis of a reconstructed CPI that gives food and energy components a zero weight. There are also related measures – inflations based on 'CPI less energy', 'CPI less food', the median CPI, and various 'trimmed mean' CPIs. The use of 'CPI less $x$' measures of core inflation is sometimes justified on the grounds that the excluded terms (supposedly) have more noise than the included terms, and the median and trimmed mean measures can be justified on robust statistics and other grounds (see, e.g., Bryan *et alia*, 1997). Core inflation measures are also sometimes obtained using regression-based methods, which can be regarded as producing optimal measures of core inflation given assumptions about the trend itself and the noise around it. These methods include moving averages, exponential (and other) smoothing methods and time series (Box-Jenkins) methods.

This paper proposes some new core inflation measures based on a recently developed approach – wavelet analysis – that is ideally suited to the estimation of core inflation. The motivation for these measures is the simple idea that if core inflation is to be interpreted as an underlying signal from an original noisy inflation process, then we should be able to estimate core inflation using a suitable signal extraction or denoising method, but we need a method that takes account of the non-stationarity (and general 'bad behavedness') of real-world inflation series. Wavelet

---

[3] For a good selection of this literature, e.g., Blinder (1997), Bryan and Cecchetti (1993, 1994), Bryan

methods are specifically designed for this type of problem and have been used with great success in many areas of applied science and engineering. Wavelet methods avoid the arbitrariness of those approaches that estimate core inflation simply by excluding certain components from the 'parent' inflation rate. But unlike conventional statistical methods of detrending data, wavelet methods do not require strong assumptions about the trend or the noise around it. Indeed, a particular strength of wavelet methods is that they have no problems dealing with time-varying behavior that incorporates non-stationarity, jumps or discontinuities, regime shifts, isolated shocks, and similar times-series behavior, such as we see in many inflation rate series. Wavelet methods are therefore ideal for signal extraction problems such as the measurement of core inflation.[4]

This paper is laid out as follows. Section 2 briefly reviews some of the existing literature on core inflation in the United States. Section 3 introduces wavelet methods. Section 4 explains their suitability for estimating core inflation and shows how they can be applied to this purpose. Section 5 explains the data, and section 6 presents our results. Section 7 concludes.

---

*et al*. (1997), Cecchetti (1997), Clark (2001), Cogley (2002), Smith (2004), and Wynne (1997).

[4] As well as the two performance criteria applied of trend tracking and predictability it is clearly important to be able to apply wavelet analysis in real-time contexts. Here we would want the wavelet measures of core inflation to update the estimates based on the arrival of new data in a reliable fashion. This is a developing area, and applications have been made on real-time problems such as processing new ethernet and ATM data using Abry-Veitch wavelet-based estimators (see Roughan *et alia*, 2000). The idea is to use a pyramidal filter bank with an algorithm that allows new incoming data be processed individually and merged with existing processed data in a way that does not require a complete recomputation of the new entire data set. Alternatively, an ad hoc approach that is computationally less efficient would be to use standard wavelet methods that would be updated on a

## 2. EXISTING MEASURES OF CORE INFLATION

*2.1. CPI-based measures of core inflation*

Many common measures of U. S. core inflation are derived from the CPI as the 'parent' inflation series. Perhaps the most widely used such measure is 'CPI inflation less food and energy' – the CPI inflation rate constructed with zero weights attached to the food and energy components – which has long been used as a measure of core inflation in the U.S. This measure is often motivated informally on the grounds – valid or otherwise – that food and energy are subject to a lot of high frequency variation (e.g., due to weather, seasonality, etc) that injects noise into CPI inflation signals. It is argued that cutting out these components eliminates this noise and so gives a better indicator of underlying (i.e., core) inflation. We can also cut out these components individually to give us two different core inflation measures – 'CPI less energy' and 'CPI less food' – that can be motivated on similar grounds, and we can cut out other components as well. These measures have the advantages that they are easily understood, timely and not subject to revision. However, the choice of excluded components is arbitrary, such measures fail to address relative price shocks

---

rolling window basis to give real time estimates of core inflation. These approaches offers promise for further developments in the estimation of real time core inflation measures.

within included components, and the excluded components may not be more volatile than included ones.[5]

A common alternative is the median CPI inflation rate, proposed by Bryan and Cecchetti (1994), Bryan *et al*. (1997), Apel and Jansson (1999), and others. The median has the advantage that it does not require the arbitrary selection or deselection of components in advance. It is also more robust than the CPI inflation rate to large shocks in individual components; its use can therefore be justified by reference to the theory of robust statistics (i.e., that in the presence of skewness and other non-normality, the median is a more efficient estimator of the population mean than the sample mean). Its use can also be justified by the economic argument that in the presence of menu costs, firms will only change prices in the face of large shocks; in such circumstances, the median would be a better estimate of the underlying inflation than the mean. However, the median excludes components experiencing relatively large price changes and when used as a measure of core inflation may therefore miss price changes that provide useful information on trend inflation (Cockerell, 1999; Clark, 2001). It has also been said that central banks may not wish to use the median as a measure of core inflation on the grounds that the median is not well nderstood by the public (Rogers, 1998; and Álvarez and Llanos, 1999).

---

[5] Most empirical studies seem to suggest that CPI less food and energy is a poor measure of core inflation (see, e.g., Cecchetti, 1997), but there appears to be little consensus on the performance of other 'CPI-less-*x*' measures of core inflation.

Closely related are the trimmed mean measures.[6] These are the means of the ordered, weighted component inflation rates, with the upper and lower tails excluded. For example, the 9% trimmed mean is the mean constructed from the ordered weighted component inflation rates, with the bottom 9% and top 9% of changes excluded. The use of these measures can also be justified by robust statistics theory and/or economic menu-cost arguments, and they too do not require arbitrary decisions on what components should be included or excluded.[7] However, trimmed measures arguably suffer from similar problems as the median.

*2.2. Regression-based measures of core inflation*

Regression methods provide a rather different approach to core inflation measurement. If we define core inflation in terms of a trend, then the estimation of core inflation requires estimation of a trend and – provided we feel comfortable with the statistical assumptions involved – regression methods provide a straightforward way to estimate this trend: we specify our underlying assumptions about the trend itself (e.g., whether it is linear, quadratic, etc.); we specify the nature of the the departures from it, which can be regarded as the equation's errors (e.g., these might be normal, autocorrelated, etc.); and we apply the appropriate regression method. These methods include simple moving averages of current and past inflation, time series methods such as Box-Jenkins ARMA and ARIMA methods,[8] and statistical smoothing methods, such as the exponentially smoothed core inflation estimator recently proposed by Cogley (2002).[9] Regression-based methods avoid the arbitrariness of including or excluding particular components, are easy to use, and can be applied in real time if we estimate regressions on a rolling basis. They also have the attraction that a suitable choice of regression method can provide us with an optimal trend – and hence an 'optimal' core inflation measure – *given* the assumptions we make. However, their disadvantage is that we have no easy way to verify what these assumptions should actually be. In practice, we can rule out the more obviously naïve assumptions (e.g. that core inflation is constant?) on a priori grounds or on the basis of suitable tests, but distinguishing between more plausible alternatives is difficult. In using such methods, we therefore end up having to make certain assumptions on faith, and the value of our results may be dependent on the validity of these untested assumptions.

---

[6] There are also other related measures. For example, there are the volatility-weighted (or Edgeworthian) measures, proposed by Dow (1994), Diewert (1995), Wynne (1997, 2001) and Vega and Wynne (2001). These are based on the idea that we should weight price changes by volatility, so giving more volatile components a lower weight and getting a more accurate aggregate price signal. Since volatility erodes the relative price signal, we volatility-adjust the weights to get a more reliable aggregate signal. These have the advantage over median and trimmed mean estimators of not throwing information away, but constructing weights can be difficult and the weights themselves can be unreliable.

[7] Most studies seem to agree that these median and trimmed means measures generally perform better than CPI less food and energy. There does not appear to be any consensus on the relative performance of median and trimmed mean measures, although one study (Smith, 2004) clearly finds that the median performs better than the trimmed-mean measures.

---

[8] There is relatively little clear evidence in the literature on the performance of these measures, because these measures are usually used in this area to provide proxies for trend inflation against which *other* measures of core inflation can be assessed. In this paper, by contrast, we are concerned with these estimators as measures of core inflation in their own right.

[9] This estimator has a number of attractive features as a measure of core inflation (e.g., it depends on a single parameter that is easily calibrated, has a nice theoretical basis in Sargent's self-confirming expectations theory, etc.), and Cogley's results suggest that it performs well.

## 3. WAVELET METHODS

A wavelet can be defined as a "waveform of effectively limited duration that has an average value of zero" (Misiti *et al*. (2000, p. 1-9)). More informally, we can think of a wavelet as a localised waveform in time-scale space. Wavelet analysis consists of taking a chosen waveform, the so-called mother wavelet, and breaking up a signal or series into shifted (i.e., in time) and scaled (i.e., compressed and extended) versions of the mother wavelet. Unlike classical wave (i.e., Fourier) analysis,[10] a wavelet does not require strong statistical assumptions and can be applied to 'difficult' signal processing problems such as the detection of underlying patterns or trends in non-stationary series, the removal of noise, and the identification of breakdown points and discontinuities. Wavelet analysis is a relatively recent development – most of the seminal work was done in the 1980s – but has proven to be immensely useful in many diverse areas such as medicine, biology, oceanography, earth studies, and fingerprint analysis. Wavelets have also been used in a variety of economic and finance applications, such as the modelling of non-stationary processes (Ramsey and Zhang, 1997) and long-memory processes (Jensen, 1999), time-series decomposition (Ramsey and Lampart, 1998), forecasting (Stevenson, 2001), scaling analysis (Gençay *et al.*, 2002), and outlier testing (Greenblatt, 1994).[11]

---

[10] Fourier analysis is limited because it presupposes a stationary signal. The traditional solution to the problem of analysing non-stationary series within the Fourier paradigm is to use short-time or windowed Fourier analysis (Gabor, 1946): we break up our series into a series of fixed-size windows and apply Fourier analysis to each window separately. However, this is a very imperfect solution (e.g., because of the arbitrariness of the size of the window), and it is much better to use wavelets instead. Wavelets are more flexible and much more powerful – not least because they enable us to obtain greater accuracy in both scale and time domains.

Wavelet analysis is particularly promising for the estimation of core inflation for several reasons. First and most obviously, wavelet analysis is tailor-made for the denoising of (or signal extraction from) non-stationary time series, and these characteristics are ideal for core inflation analysis: the process of obtaining a core inflation rate is essentially a form of signal extraction, and inflation series are typically non-stationary. Hence, wavelet analysis has the advantages that it avoids strong statistical assumptions (unlike, say, regression-based estimates of core inflation) and has no problems handling the non-stationary behavior of inflation arising from shifts in monetary policy, oil price shocks, and the like.

Secondly, because wavelet analysis can lead to core inflation series that have fundamentally different 'shapes' over time, we can use it to select a core inflation series that reflects the reasons why we might want a measure of core inflation in the first place, i.e., we can choose a core inflation series whose 'shape' helps us with the problem at hand. Some wavelet specifications give rise to core inflation series whose times-series plots are 'smooth', whilst others give rise to plots

---

[11] For more on these and other economic applications, see also Gençay *et al.* (2002) and Schleicher

that are 'pointed' or exhibit 'plateaux'. Each of these shapes can satisfy a particular purpose. For example, if we want a core inflation measure that reflects an underlying trend, we would presumably want our core inflation curve to be smooth. However, there may be circumstances where we prefer one of the other shapes. For instance, if we were more interested in capturing the key inflation turning points, we might prefer a 'pointed' shape because this shape highlights the turning points in which we are primarily interested. Alternatively, we might be interested in distinguishing between shifts in core inflation, and in this case the 'plateaued' [sic] shaped would be more helpful: each flat plateau represents a particular period of 'stable' core inflation, and shifts from one plateau to another represent a jump from one core inflation rate 'sub-regime' to another. Thus, wavelets can be used to produce core inflation measures that are tailor-made to the context in which we want to use them.[12, 13]

---

(2002).

[12] Wavelet methods also have another, albeit mixed, blessing. From a filtering point of view, wavelet analysis can be regarded as a highly efficient form of two-sided filtering with time-varying coefficients, and have the standard advantages of two-sided over one-sided filtering. (For more on these, see, e.g., Gençay *et alia*, 2002.) So, for instance, in estimating unobservable variables such as core inflation over a specific time period, wavelet methods can incorporate data from both before and after that time period, and so produce estimates that are superior (from a filtering perspective) than those that can be obtained using only one-sided filters (in which estimates of the current period's core inflation make use of data available in this period and do not make use of later data that becomes available). This is an advantage in contexts where we might be interested in measuring historical core inflation rates. Nontheless, it is a (potential) disadvantage when operating in situations where we wish to estimate core inflation using only past data for each period concerned, i.e., in many real-time contexts. However, as we have seen in note 4 above, this is not an insurmountable barrier by any means.

[13] One potential problem with wavelets applied to core inflation is that they are essentially 'black box' methods whose inner workings are invisible to the public. By contrast, some of the tradition methods like 'CPI ex' and trimmed mean methods have some intuitive explanation that can be conveyed to the general public. However, this kind of communicability to the public is only one characteristic of a desired core inflation method, and this disadvantage of wavelet methods needs to be weighed against their attractions. In any case, it is not always the case that a core inflation measure needs to be

To apply wavelets, we first select a particular wavelet type, and there are many types to choose from; these have different properties and are useful for different applications.[14] The simplest (and also first known) wavelet is the Haar wavelet, which is a straightforward step function. A second is the Mexican hat wavelet with its distinctive 'Mexican hat' shape. Other popular wavelets are the family of Daubechies wavelets. These are distinguished from each other by their order number – Daubechies 1, Daubechies 2, and so on – and include the Haar wavelet as a special case when the order is 1. Similar to these are the symlets – symlet 1, symlet 2, etc. – which are nearly symmetrical relatives of the Daubechies family.

Some of these wavelet shapes are illustrated in Figure 1. This Figure shows the discontinuous Haar or step-function wavelet, the unmistakeable Mexican hat wavelet, a Daubechies 4 wavelet showing the asymmetry of this wavelet family, and a near-symmetrical Symlet 2 wavelet.

**Insert Figure 1 here**

---

'explained' to the public: for example, a central bank might use wavelet methods to provide inflation indicators for its own internal analysis. One might also argue that the public has little real understanding of index-number issues in the first place.

[14] There exists a considerable technical literature on the different wavelets and their properties. Perhaps the main criteria useful in evaluating which wavelets suit which applications are: the support, which affects speed of convergence as the frequency gets large; the degree of symmetry, given that symmetry is an advantage in image processing; the number of vanishing moments, which affects compression analysis; and regularity, which is useful for getting nice features such as smoothness in reconstructed signals. For more on the different wavelets and their properties, see Daubechies (1992) and Misiti *et al.* (2000).

Having selected our wavelet type, we then apply a suitable algorithm. In most practical cases this would be Mallat's discrete wavelet transform, which is a fast and efficient way to estimate the wavelet coefficients (Mallat, 1989). Applying such an algorithm decomposes our original series into two series – an 'approximation' series, which highlights the high-scale, low-frequency components (or underlying pattern or trend) of the original series, and a 'details' series, which highlights the low-scale, high-frequency components (or noise) of the original series. If we wish, we can then take the approximation series and filter it again: this gives us a level 2 approximation. We can repeat the process as often as we wish: filtering the level 2 approximation gives us a level 3 approximation, and so on. Each time we apply the filter (or increase the approximation level) we remove more of whatever high-frequency, low-scale components still exist in our approximation series, and so obtain a 'cleaner' high-scale, low-frequency underlying pattern or trend. We can therefore think of the application of wavelets as involving both a choice of wavelet form and a choice of approximation level: decomposing the series once is a level-1 analysis, decomposing twice is a level-2 analysis, and so on.

If we apply this method to a given inflation series, then the suitably 'cleaned' approximation series gives us our core inflation, and the original, given, inflation series is our 'parent' inflation. Put another way, we take our 'parent' series, and put it through a wavelet analysis to obtain our core inflation series, and the choice of wavelet type and approximation level determine how the core inflation series turns out.

# 4. USING WAVELETS TO ESTIMATE CORE INFLATION[15]

As we have already mentioned, wavelet analysis involves the application of successive approximations to remove more and more high frequency detail, or noise, and so help reveal an underlying signal (or shape). However, if we apply too many approximations, we also remove parts of the signal itself. A balance therefore has to be struck so that we don't lose the signal along with the noise. This means that we need some guidelines to help draw a reasonable balance, and in this paper we make use of two particular sets of guidelines:

- *Normality of 'details'*: There is a heuristic argument that we should stop approximating before the details have become normal. More precisely, if the details at level $i$ are normal, then the $i^{th}$ level approximation merely removed 'random' noise, and this would indicate that the $i^{th}$ level approximation was unnecessary: i.e., no more than $i$-1 levels are needed.

- *Minimum (or, more generally, low) entropy*: Applying information theory, the optimal number of levels is the one with the lowest entropy. More generally, any reasonable choice for the number of levels should have a relatively low entropy.

---

[15] Since our purpose is to introduce wavelet methods to the estimation of core inflation, there is little point in using some of the more sophisticated tools in the wavelet armoury such as wavelet thresholding methods. In any case, experimentation with thresholding methods indicates that these make only insignificant changes to the core inflation series. Such refinements were therefore not used in the results reported here.

These considerations suggest the following approach to the selection of plausible wavelet-based measures of core inflation:

- We start with a reasonable universe of plausible wavelet types,[16] where the initial choice of plausible wavelets might be guided by the context we are working in: so, for example, if we are seeking a 'smooth' core inflation series, we might restrict ourselves to wavelets that generate a smooth core inflation series, etc. For each type of wavelet, we consider a reasonable number of approximation levels.[17]

- For each wavelet type and approximation level considered, we estimate the Jarque-Bera (JB) probability values for the details estimated at that level of approximation, and we note those levels (if any) with reasonable prob-values (e.g., higher than 1%). If we cannot find any levels with reasonable JB prob-values, then we eliminate that wavelet from consideration.

- For those wavelets that remain, we estimate their entropies and note how entropy values change with the level of analysis. This entropy analysis should give us a range of permissible levels of approximation, as judged by entropy criteria.[18]

---

[16] The wavelets considered in this study were those that could be calculated by MATLAB's Wavelet Toolbox, which covers the most popular wavelets.

[17] We considered up to 10 such levels, which should be more than adequate in most contexts.

[18] We looked at Shannon and log energy entropies (see, e.g., Misiti *et al.*, 2000), and in every case, we found that the entropies were lowest for relatively low levels of approximation. This indicates the need to take a relatively low number of approximation levels: certainly no more than 5 in any case considered, and in many cases ideally lower.

- We now eliminate all combinations of wavelet and approximation level that do not satisfy *both* our JB prob-value and our entropy-permissible criteria.
- We can cut down further by doing a casual inspection of our candidate core inflation series and eliminating those that look very similar to each other.[19] The remaining series – in our case, just 6 – are our plausible core inflation series. [20]

## 5. DATA

Our 'parent' inflation series is the year-on-year rate of change of seasonally-adjusted CPI (all items inclusive), observed on a monthly frequency.[21] The choice of CPI can be justified by its importance for monetary policy purposes, by its significance as a headline inflation index, by the fact that it is perhaps the most commonly used 'parent' series in core inflation studies (and this in turn allows for meaningful comparisons with existing studies), and by the fact that it allows for a wide variety of different core inflation series.[22] We use seasonally adjusted series to minimize seasonal noise, and because most other studies also use seasonally adjusted data.

---

[19] In the present case, a little over half the series were judged to be very close to other series, and were therefore eliminated as having relatively little value added.

[20] Plots and calculations were carried out using Eviews and MATLAB, including MATLAB's Wavelet Toolbox. The MATLAB programs specially written for this paper are available on request.

[21] Alternatively, we might have worked with annualised monthly rates of change. Such series are considerably more volatile (and therefore arguably a more interesting subject for a denoising analysis), but we prefer to report results based on the more conventional year-on-year change to facilitate comparison with most other studies, which also work with year-on-year changes.

We use a long data period from February 1967 to January 2002 to make maximum use of available data. The length of our data set is constrained on the one hand by the fact that some of our series only start in February 1967, and on the other hand by the fact that our trimmed mean series end in January 2002. This data set encompasses 421 monthly observations over an interesting period that encompasses significant shifts in monetary policy and the behavior of inflation.

## 6. EMPIRICAL RESULTS

We look at a variety of CPI-based, regression-based and wavelet-based measures of core inflation:

- The CPI-based measures are: (1) CPI inflation less food and energy, (2) CPI inflation less energy, (3) CPI inflation less food,[23] (4) median CPI inflation,[24]

---

[22] Naturally, we recognise that the CPI has its problems. Most particularly, it has undergone major changes from time to time and these may impact on our results. However, such problems arise with all studies that use the CPI, and there is little we can realistically do about them here. Instead, we simply take the CPI as a given 'parent' inflation series for reasons explained in the text, and then focus on alternative measures of core inflation that can be derived from it.

[23] The series used are 'CPILFESL: Consumer price index for all urban consumers: all items less food & energy, seasonally adjusted', 'CPILEGSL: consumer price index for all urban consumers: all items less energy, seasonally adjusted, and 'CPIULFSL: Consumer price index for all urban consumers: all items less food, seasonally adjusted', which are all available on the FREDII website. Note that all inflation rates were constructed as the difference in the logs of relevant price indices.

[24] The series used here is taken from the Cleveland Fed website at http://www.clevelandfed.org/Research/Data/mcpi.txt.

- (5) 9% trimmed mean CPI inflation, and (6) 18% trimmed mean CPI inflation.[25]

- The regression-based methods are: (7) a 'long MA', a simple average of the current and previous 36 observations of the CPI inflation rate; (8) a 'short MA', a simple average of the current and previous 18 observations of the CPI inflation rate;[26] (9) Cogley's exponentially smoothed core inflation measure, with its single parameter set at 0.125/3 per month in line with his recommended value of 0.125 per quarter;[27] and (10) an ARMA(1,1) fitted to the parent CPI inflation rate.

- Using the selection approach outlined earlier (see above, pp. 18-19), we ended up choosing the following six wavelets for further analysis: (11) a Daubechies 10 wavelet obtained at a level 4 approximation; and (12) a symlet 5 wavelet at a level 5 approximation; (13) a Daubechies 2 wavelet at a level 3 approximation; (14) a Daubechies 3 wavelet at a level 5 approximation; (15) a Haar wavelet at a level 2 approximation; and (12) a symlet 1 wavelet at a level 2 approximation. The first two, second two and final two sets of wavelets provide good illustrations of 'smooth', 'pointed' and 'plateaued' patterns, respectively, and were chosen for their illustrative properties without regard to their potential test performance.

---

[25] These latter two series are constructed using the CPI components and relative weights given by Smith (2004), available in the JMCB data archive at http://webmail.econ.ohio-state.edu/john/IndexDataArchive.php.

[26] Based on the series 'CPIAUCSL: Consumer price index for all urban consumers: all items, seasonally adjusted', available on the FREDII website.

*Plots of core inflation*

Before carrying out more formal comparisons, it is good practice to plot the various core inflation series and look for any outstanding features. Plots of our series are given in Figures 2-6: Figure 2 gives plots of the 'CPI-ex' core inflation measures, Figure 3 gives plots of the regression-based measures and Figures 4-6 give plots of our chosen wavelets. In each case, we also plot the 'parent' CPI inflation rate for comparison. We separate out the wavelet plots so that Figure 4 gives plots of the two 'smooth' wavelet measures, Figure 5 gives plots of the 'pointed' wavelet measures, and Figure 6 gives plots of the 'plateaued' wavelet measures. A comparison of these Figures indicates that there are some striking differences between different core inflation series.

**Insert Figures 2-6 here**

*Evaluation criteria*

The performance of core inflation measures can be evaluated using a number of criteria. As outlined earlier we have two performance criteria for our candidate measures that are based on alternative definitions of core inflation: ability to track a trend and predict future inflation.

---

[27] Because the MA and Cogley measures involve measures of core inflation that involves long lag functions of the parent series, these particular measures were estimated using a parent inflation data set going back to February 1962.

Beginning with trend based evaluation, we might expect a good core inflation series to have the same mean as its 'parent' inflation series (see, e.g., Clark, 2001, Table 2). Moreover, a good core inflation measure should also have a lower variance than its parent series, and this suggests that the core series should pass a test of the hypothesis that its variance is less than that of its parent series. A trend-based notion of core inflation should also lead us to expect that a core inflation series would exhibit fewer turning points than its parent series. In addition, we might expect a good measure of core inflation to be cointegrated with its parent inflation series (e.g., as in Freeman, 1998, Marques *et al.*, 2000), and we might also expect the difference between these two series to be stationary.

If we think in terms of core inflation as a predictor of future inflation, then we are looking for evaluation criteria that address the 'closeness' between core inflation now and actual inflation later. This suggests that we might use the variance of the difference between current core inflation and later actual inflation as an evaluation criterion: the lower the variance, the better the fit. We might also expect core inflation now and actual inflation later to be cointegrated, and the prediction error to be stationary.

We can also examine whether a candidate core inflation series passes a recent inflation-prediction test suggested by Cogley (2002). If $\pi_t^c$ is a measure of

(inflation predicting) core inflation, then for any reasonable horizon *H*, the regression:

(1) $$\pi_{t+H} - \pi_t = \alpha + \beta \, (\pi_t - \pi_t^c) + u_{t+H}$$

should satisfy the predictions $\alpha = 0$ and $\beta = -1$. These restrictions reflect the expectations that a good core inflation series should predict future changes in inflation by the right magnitude. So for example, if the slope coefficient is less (greater) than one in an absolute sense it suggests that the measure of core inflation is overpredicting (underpredicting) the magnitude of subsequent changes in inflation. We would also expect the regression to exhibit a good overall fit (e.g., a high $R^2$).

*Results*[28]

Table 1 reports summary results for the means, volatilities and the numbers of turning-points of each core inflation series. These are presented as the relevant core inflation parameters divided by their parent inflation counterparts:

- *Means*: Most series have mean ratios indicating that their means are close (or even identical) to those of the parent CPI series. The accuracy of the wavelet measures in particular is striking. The only series that perform poorly by this

---

[28] It has been argued that a good core inflation measure should also remove any seasonality in the original inflation series (see, e.g., Cecchetti, 1997). However, tests of seasonality are not reported in

criterion are the trimmed mean, short MA and exponentially smoothed measures. These are respectively 8.9%, 6.1% and 7.1% away from their target of 1 whereas the other ratios are generally within 1% of the target ratio.

- *Variances*: The CPI-ex measures have very variable variance ratios, ranging from 0.386 (18% trimmed mean) to 0.948 (CPI less food). The regression-based ratios tend to be lower and vary from 0.295 (exponentially smoothed) to 0.509 (short MA). The wavelet ratios tend to be in-between the others and vary from 0.420 (db3, level 5) to 0.663 (Haar, level 2).

- *Numbers of turning points*: The CPI-based measures clearly perform badly by this criterion, as their turning point ratios all exceed 1. The regression-based measures perform better with turning point ratios varying from 0.542 to 0.792, and the wavelet-based measures perform much better still with ratios varying from 0.064 to 0.441.

**Insert Table 1 here**

These results indicate that CPI based measures tend to perform poorly across one or more criteria – the trimmed mean, short MA and exponentially smoothed measures do badly by the mean criterion, CPI-less-food does badly by the variance criterion, and the CPI-ex measures all do badly by the turning point

---

this paper partly because we use seasonally adjusted data, and partly because all our core inflation series easily passed these tests anyway.

criterion. The performance of the regression-based measures is mixed, and that of the wavelet-based measures is pretty good.[29]

Results for the tests of cointegration between core and parent inflation series are presented in Table 2. These give a clear picture: all the results reported indicate that the relevant pairs of series are cointegrated.

**Insert Table 2**

Results for tests of the stationarity of the difference between core and parent inflation series are presented in Table 3. For every series, the null hypothesis of a unit root is easily rejected: all series pass with flying colors. Thus, these first two tests give us no reason to prefer any one measure of core inflation to any others.

**Insert Table 3**

We now turn to the inflation-prediction tests, and Table 4 gives the variances of core inflation prediction errors applied to an 18-month forecast horizon.[30] If we rank them by their lowest-first variance rankings, the CPI-based measures have rankings varying from 6 to 16 (with most of them at or near the bottom), the regression-based measures have rankings of 7 to 13, and the wavelets have rankings from 1 to 10 for the Haar wavelet (and include the top 5 rankings). Note, too, that the performance of the wavelet measures improves even further if we exclude the 'rogue' Haar wavelet: as Figure 1 shows, this is a fairly 'primitive' wavelet anyway. Thus, a fairly clear pecking order emerges: the wavelets generally perform best by this criterion, and the CPI measures generally perform worst.

---

[29] As an aside, note that we have compared each core inflation series to the *actual* parent series. However, it is quite common in core inflation studies to compare core inflation series to some *assumed* trend series. For example, some authors look at whether the core inflation improves upon simple autoregressions of the actual inflation rate, whilst others take a moving-average trend as a proxy for core inflation and then assess the candidate core against the proxy (e.g., via lowest variance or RMSE criteria). However, such methods are questionable because they are based on the assumption that we already know how to model the trend correctly, and this begs the very point at issue, i.e., that the whole purpose of the investigation is to identify the 'best' trend, given that we are working here with a trend-based interpretation of core inflation. We can also put this point a different way: if we think that we already know what the 'best' trend is, then the investigation is redundant, because we already have the answer; and, on the other hand, if we don't know what the 'best' trend is, then we cannot use some arbitrary rule as a proxy for it. Any significant difference between the candidate core inflation series and the proxy is then uninterpretable without further information, because we could not dismiss the possibility that it is the proxy rather than the core inflation measure that is 'wrong'.

**Insert Table 4**

Results for tests of cointegration between core and future parent inflation series are presented in Table 5. These indicate that the hypothesis of no cointegration is almost always highly implausible.[31]

**Insert Table 5**

Results for tests of the stationarity of the core inflation prediction error are presented in Table 6, and indicate very clearly that the prediction errors are stationary.

**Insert Table 6**

---

[30] We also replicated the prediction results for a 12-month and 24-month forecast horizon. The results for a 12-month forecast horizon tend to produce very similar rankings; for a 24-month horizon the superiority of the wavelets-based measures over the regression-based ones is still discernable but less marked. These results for prediction errors over alternative forecast horizons are available on request.

[31] The results reported in this Table and the next are extremely robust if we change the horizon periods to 12 or 24. Those reported in Table 7 are broadly robust (i.e., do not affect the main conclusions) in the face of such changes.

Table 7 gives the prediction results based on estimates of (1). The Table also reports the individual coefficients of the regression and their standard errors, and these can be used to carry out $t$-tests of the two predictions $\alpha = 0$ and $\beta = -1$ separately considered, and to carry out an $F$-test of them jointly. We find that the CPI-based measures fail these tests, the regression-based measures give a mixed performance, and the wavelet-based measures (usually) perform well. The Table also reports the different measures' $R^2$ rankings. The CPI-based measures have the 6 lowest rankings, the regression-based measures have rankings from 4 to 8, and wavelet-based measures have rankings from 1 to 10 (and include the top 3 performers). A similar pecking order therefore emerges: the wavelets generally perform best, and the CPI-based measures perform worst, and the regression-based measures (usually) perform somewhere between.

**Insert Table 7 here**

If we take all these results together, we can also say something about the relative performance of individual series in each of these groups:

- *CPI-based measures*: There is not a great deal to choose from between 'CPI less $x$' and median CPI measures. Amongst the CPI-based measures, the trimmed mean measures perform worst in the ratio of means results in Table 1, but in contrast outperform the other CPI measures in the inflation-prediction results of Tables 4 to 7.

- *Regression-based measures*: There is also little to choose from between most of these measures, except to point out that the exponentially smoothed measure performs worst by the ratio of means results in Table 1 but best in the inflation-prediction results in Tables 4 to 7.

- *Wavelet-based measures*: Amongst the wavelets, the worst performer is clearly the Haar measure: this measure is not only the worst wavelet in each ranking, but its ranking is always well out of the range of the others. Eliminating this measure would therefore significantly improve the wavelet rankings overall. The wavelet results overall also suggest that the 'smooth' wavelets generally do better than the 'pointed' ones, which in turn tend to do better than the 'plateaued' ones. Thus, our best performers appear to be the 'smooth' ones, the db10 (level 4) and sym5 (level 5), and this makes them very natural measures of core inflation if we want a 'smooth' core inflation series. It is also interesting to note that these two measures have especially low turning-point ratios – suggesting the not-unreasonable conclusion that a good measure of (trend) core inflation should have relatively few turning points.

## 7. CONCLUSIONS

This paper has suggested using wavelet methods to estimate core inflation. Wavelets are ideally suited to denoising non-stationary time series, and the problem of estimating core inflation is essentially one of denoising a 'badly behaved' time series. An additional attraction of using wavelets for this purpose is that different wavelets lead to core inflation series of different shapes, so we can choose a wavelet that produces the shape we think most appropriate to the precise problem at hand, i.e., one which addresses the question of why we want to measure core inflation in the first place. The paper explains how wavelets might be applied to this purpose, and sets out a wavelet selection procedure that will generate a core inflation series from some initial 'parent' inflation rate. The paper goes on to compare wavelet-based measures of core inflation against a number of existing measures, and results indicate that the wavelet-based measures generally do better, and often much better, than earlier core inflation measures.[32] The performance of the wavelet measures is particularly impressive, and not least because the wavelets were chosen merely for illustrative purposes and no effort was made to find an 'optimal' wavelet measure. There is therefore every reason to believe that an 'optimal' wavelet would perform even better.[33]

---

[32] The choice between different wavelets comes down, in part, to which of these types of trends is most plausible for the application at hand. However, from the point of view of identifying core inflation as trend inflation, the 'smooth' trends are presumably more plausible in this context than the other ones. Interestingly, the two individual wavelets that dominate the performance evaluation are both examples of wavelets that produce 'smooth' core inflation series.

[33] Two obvious extensions to our work are therefore as follows. (1) How more sophisticated wavelet methods might yield even better measures of core inflation. (2) How wavelet methods of core inflation might be implemented in a real-time context using only information currently available.

Wynne, M. A. (1999) "Core inflation: A review of some conceptual issues." European Central Bank *Working Paper Number 5.5.*

Wynne, M. A. (2001) "The Edgeworth index as a measure of core inflation." Mimeo. Federal Reserve Bank of Dallas.
35

# FIGURES

## FIGURE 1: Illustrative Wavelets

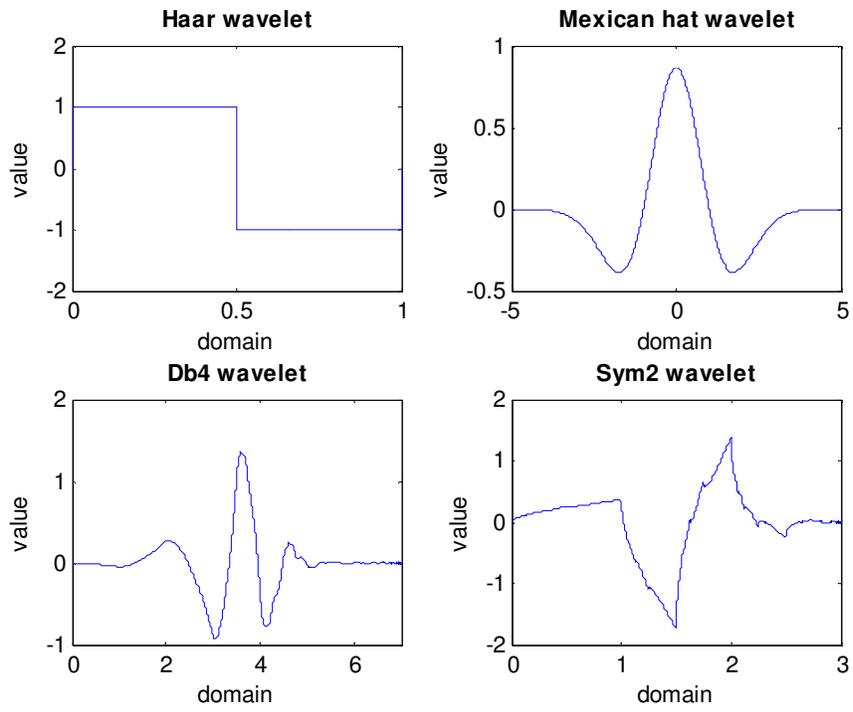

**FIGURE 2: CPI-Based Measures of Core Inflation**

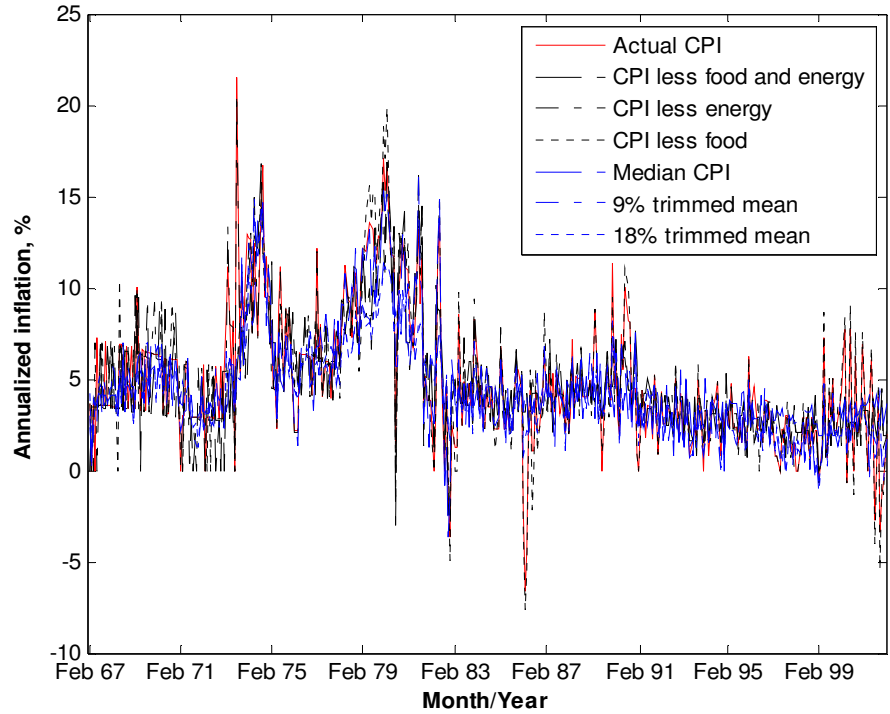

Notes: Based on 421 monthly observations spanning February 1967 to January 2002.

**FIGURE 3: Regression-Based Measures of Core Inflation**

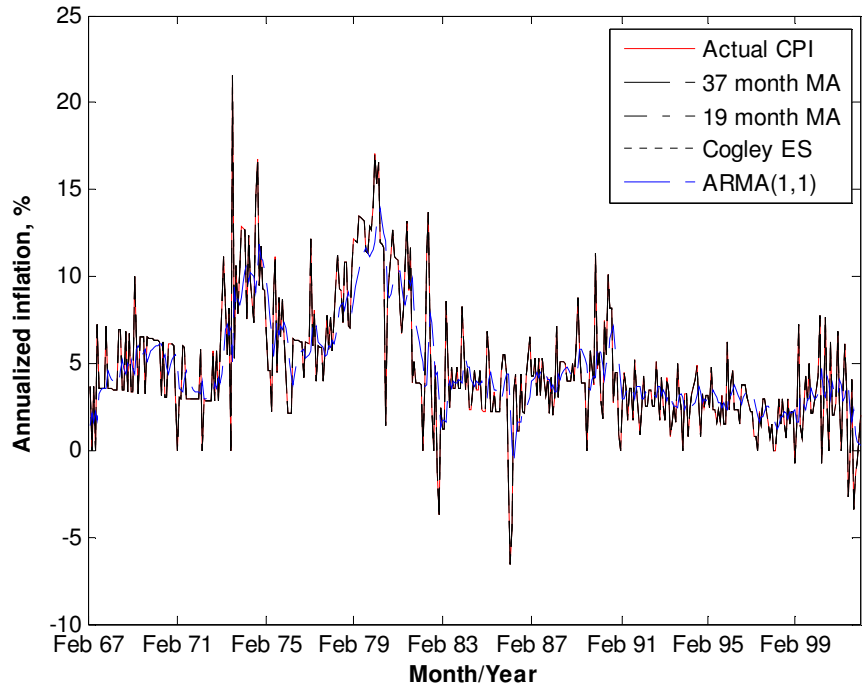

Notes: Based on 421 monthly observations spanning February 1967 to January 2002.

**FIGURE 4: 'Smooth' Wavelet-Based Measures of Core Inflation**

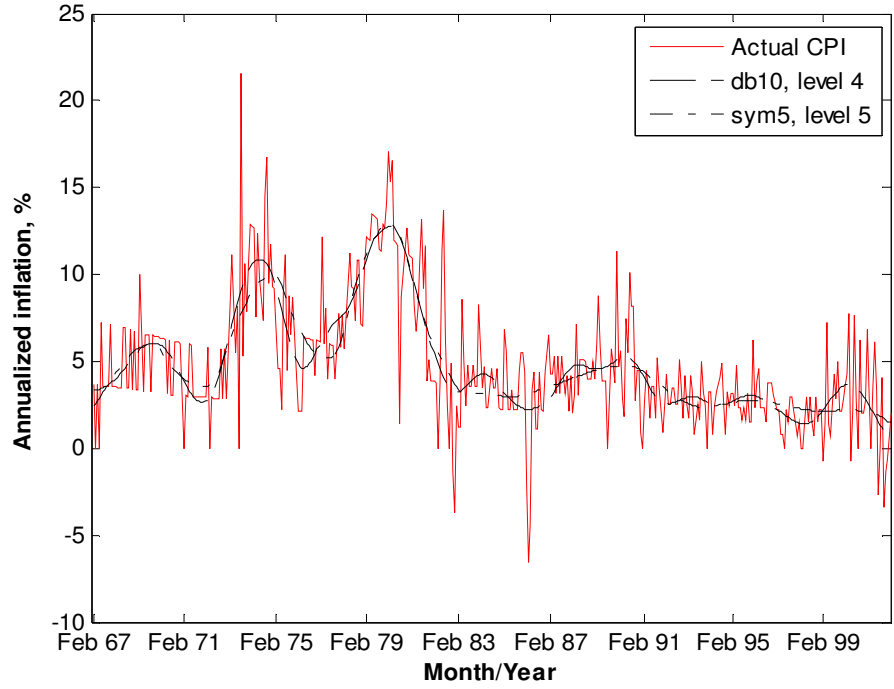

Notes: Based on 421 monthly observations spanning February 1967 to January 2002.

**FIGURE 5: 'Pointed' Wavelet-Based Measures of Core Inflation**

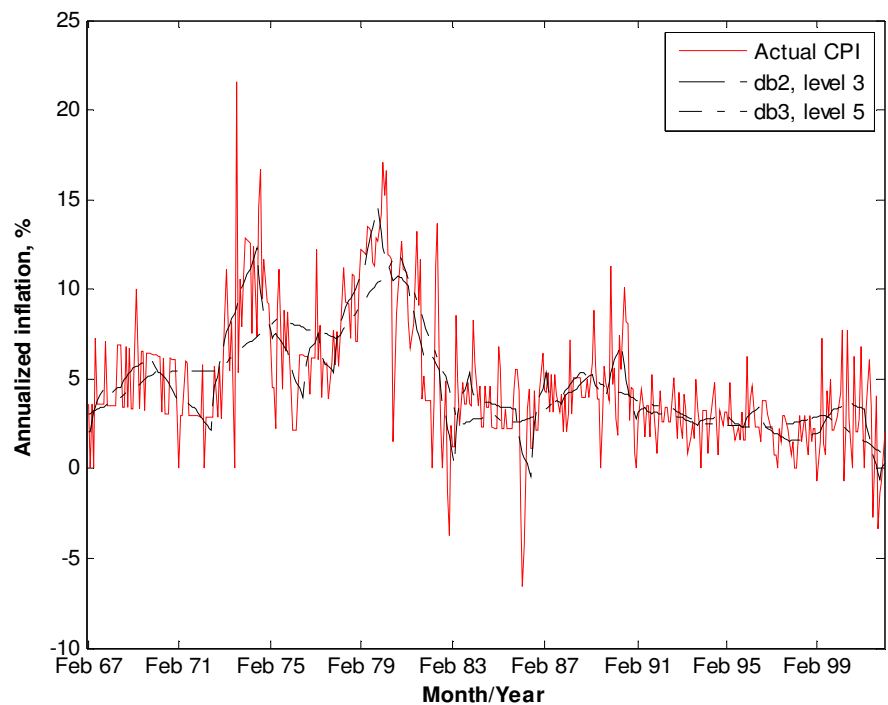

Notes: Based on 421 monthly observations spanning February 1967 to January 2002.

**FIGURE 6: 'Plateaued' Wavelet-Based Measures of Core Inflation**

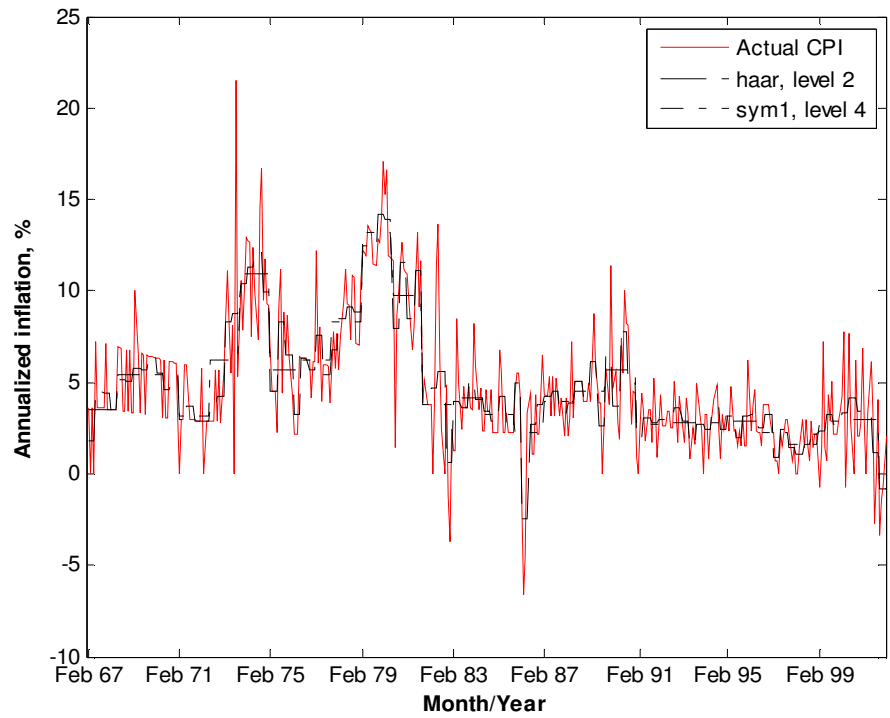

Notes: Based on 421 monthly observations spanning February 1967 to January 2002.

# TABLES

## Table 1: Summary Results for Core Inflation Measures

| Core inflation measure | Ratio of means | Ratio of variances | Ratio of turning points |
| --- | --- | --- | --- |
| *CPI-based measures* | | | |
| CPI less food and energy | 1.015 | 0.635 | 1.114 |
| CPI less energy | 1.010 | 0.643 | 1.076 |
| CPI less food | 1.003 | 0.948 | 1.047 |
| median CPI | 1.011 | 0.560 | 1.140 |
| 9% trimmed mean | 0.890 | 0.468 | 1.106 |
| 18% trimmed mean | 0.890 | 0.386 | 1.110 |
| *Regression-based measures* | | | |
| long moving average | 1.004 | 0.381 | 0.602 |
| short moving average | 1.061 | 0.509 | 0.542 |
| Exp. Smooth | 0.929 | 0.295 | 0.593 |
| ARMA | 1.000 | 0.444 | 0.792 |
| *Wavelet-based measures* | | | |
| db10 (level 4) | 1.011 | 0.526 | 0.072 |
| sym5 (level 5) | 0.996 | 0.498 | 0.064 |
| db2 (level 3) | 0.999 | 0.560 | 0.271 |
| db3 (level 5) | 0.995 | 0.420 | 0.106 |
| Haar (level 2) | 1.000 | 0.663 | 0.441 |
| sym1 (level 4) | 1.000 | 0.545 | 0.110 |

*Notes*: Ratio of means is mean core inflation divided by mean CPI inflation; ratio of variances is variance of core inflation divided by variance of CPI inflation; ratio of turning points is number of core inflation turning points divided by number of CPI inflation turning points. Based on 421 observations from 67:2 to 02:1.

**Table 2: Test Results for Cointegration Between Core Inflation and CPI Inflation Series**

| Core inflation | Prob-value of hypothesis of no cointegration | |
| --- | --- | --- |
| | *No deterministic trend* | *With deterministic trend* |
| *CPI-based measures* | | |
| CPI less food and energy | 0.000 | 0.000 |
| CPI less energy | 0.000 | 0.000 |
| CPI less food | 0.000 | 0.000 |
| median CPI | 0.000 | 0.000 |
| 9% trimmed mean | 0.000 | 0.000 |
| 18% trimmed mean | 0.000 | 0.000 |
| *Regression-based measures* | | |
| long moving average | 0.000 | 0.000 |
| short moving average | 0.0001 | 0.000 |
| exp. smooth | 0.0005 | 0.005 |
| ARMA | NA | NA |
| *Wavelet-based measures* | | |
| db10 (level 4) | 0.0001 | 0.0001 |
| sym5 (level 5) | 0.0001 | 0.000 |
| db2 (level 3) | 0.0001 | 0.0001 |
| db3 (level 5) | 0.000 | 0.000 |
| Haar (level 2) | 0.0001 | 0.0001 |
| sym1 (level 4) | 0.0001 | 0.000 |

*Notes*: Results are based on 415 observations over 67:7 to 02:1. Results are for the Johansen cointegration test and we obtained using Eviews 5. 'NA' indicates that results were not available due to a near singular matrix.

**Table 3: Stationarity Test Results for Difference between Core Inflation and CPI Inflation Series**

| Core inflation | Aug. Dickey-Fuller statistic | |
| --- | --- | --- |
| | No intercept | With intercept |
| *CPI-based measures* | | |
| CPI less food and energy | -11.616** | -11.611** |
| CPI less energy | -18.789** | -18.775** |
| CPI less food | -13.075** | -13.059** |
| median CPI | -17.047** | -17.033** |
| 9% trimmed mean | -10.653** | -16.194** |
| 18% trimmed mean | -10.423** | -11.009** |
| *Regression-based measures* | | |
| long moving average | -6.379** | -6.372** |
| short moving average | -9.532** | -9.608** |
| exp. Smooth | -7.126** | -7.211** |
| ARMA | -18.890** | -18.867** |
| *Wavelet-based measures* | | |
| db10 (level 4) | -18.113** | -18.091** |
| sym5 (level 5) | -16.611** | -16.593** |
| db2 (level 3) | -12.788** | -12.773** |
| db3 (level 5) | -14.287** | -14.272** |
| Haar (level 2) | -14.489** | -14.471** |
| sym1 (level 4) | -18.086** | -18.065** |

*Notes*: ** = significant at 1% level. Test statistic is distributed as a *t*. Results are based on 419 observations over 67:3 to 02:1.

**Table 4: Variances of Core Inflation Prediction Errors**

| Core inflation | Variance | Ranking |
|---|---|---|
| *CPI-based measures* | | |
| CPI less food and energy | 17.596 | 15 |
| CPI less energy | 16.011 | 14 |
| CPI less food | 19.422 | 16 |
| median CPI | 15.303 | 12 |
| 9% trimmed mean | 13.949 | 9 |
| 18% trimmed mean | 13.103 | 6 |
| *Regression-based measures* | | |
| long moving average | 14.622 | 11 |
| short moving average | 15.811 | 13 |
| exp. Smooth | 13.298 | 7 |
| ARMA | 13.646 | 8 |
| *Wavelet-based measures* | | |
| db10 (level 4) | 12.675 | 4 |
| sym5 (level 5) | 12.156 | 2 |
| db2 (level 3) | 12.773 | 5 |
| db3 (level 5) | 11.034 | 1 |
| Haar (level 2) | 14.289 | 10 |
| sym1 (level 4) | 12.666 | 3 |

*Notes*: Results are based on 402 observations over 68:8 to 02:1, with an assumed forecast horizon of $H$=18 months. The prediction error is the difference between core inflation and actual CPI inflation at a horizon 18 months ahead. Ranking is by variance: the lower the variance, the higher the ranking.

**Table 5: Test Results for Cointegration Between Core Inflation and Future CPI Inflation Series**

| Core inflation | Prob-value of hypothesis of no cointegration | |
|---|---|---|
| | *No deterministic trend* | *With deterministic trend* |
| *CPI-based measures* | | |
| CPI less food and energy | 0.0001 | 0.0001 |
| CPI less energy | 0.0000 | 0.0000 |
| CPI less food | 0.0003 | 0.0002 |
| median CPI | 0.0000 | 0.0000 |
| 9% trimmed mean | 0.0004 | 0.0011 |
| 18% trimmed mean | 0.0003 | 0.0014 |
| *Regression-based measures* | | |
| long moving average | 0.0027 | 0.0306 |
| short moving average | 0.0003 | 0.0037 |
| exp. Smooth | 0.0006 | 0.0094 |
| ARMA | 0.0002 | 0.0012 |
| *Wavelet-based measures* | | |
| db10 (level 4) | 0.0001 | 0.0000 |
| sym5 (level 5) | 0.0000 | 0.0000 |
| db2 (level 3) | 0.0000 | 0.0001 |
| db3 (level 5) | 0.0000 | 0.0000 |
| Haar (level 2) | 0.0000 | 0.0002 |
| sym1 (level 4) | 0.0000 | 0.0000 |

*Notes*: Results are based on 397 observations over 67:7 to 00:4 with an assumed horizon of 18 months. Results are for the Johansen cointegration test and were obtained using Eviews 5.

**Table 6: Stationarity Test Results for Core Inflation Prediction Error**

| Core inflation | Aug. Dickey-Fuller statistic | |
| --- | --- | --- |
| | No intercept | With intercept |
| *CPI-based measures* | | |
| CPI less food and energy | -7.485** | -7.483** |
| CPI less energy | -8.049** | -8.045** |
| CPI less food | -4.222** | -4.225** |
| median CPI | -7.542** | -7.535** |
| 9% trimmed mean | -4.213** | -4.232** |
| 18% trimmed mean | -5.640** | -5.686** |
| *Regression-based measures* | | |
| long moving average | -4.776** | -4.774** |
| short moving average | -4.599** | -4.629** |
| exp. Smooth | -5.061** | -5.063** |
| ARMA | -5.144** | -5.140** |
| *Wavelet-based measures* | | |
| db10 (level 4) | -5.322** | -5.322** |
| sym5 (level 5) | -5.473** | -5.471** |
| db2 (level 3) | -5.466** | -5.466** |
| db3 (level 5) | -5.854** | -5.853** |
| Haar (level 2) | -5.780** | -5.780** |
| sym1 (level 4) | -5.587** | -5.845** |

*Notes*: Results are based on 418 observations over 67:4 to 02:1 with an assumed forecast horizon of *H*=18 months. The prediction error is the difference between core inflation and future actual CPI inflation.

**Table 7: Prediction-based Test Results for Core Inflation Measures**

| Core inflation measure | Intercept | Slope | F-test prob | $R^2$ | $R^2$ ranking |
|---|---|---|---|---|---|
| *CPI-based measures* | | | | | |
| CPI less food and energy | -0.121 (0.371) | -0.519 (0.064) | 0.000 | 0.352 | 14 |
| CPI less energy | -0.106 (0.338) | -0.464 (0.081) | 0.000 | 0.305 | 15 |
| CPI less food | -0.099 (0.375) | -0.353 (0.088) | 0.000 | 0.275 | 16 |
| median CPI | -0.074 (0.058) | -0.685 (0.069) | 0.000 | 0.393 | 13 |
| 9% trimmed mean | -0.393 (0.362) | -0.875 (0.070) | 0.090 | 0.456 | 12 |
| 18% trimmed mean | 0.416 (0.349) | -0.887 (0.065) | 0.084 | 0.483 | 11 |
| *Regression-based measures* | | | | | |
| long moving average | -0.085 (0.410) | -0.872 (0.053) | 0.057 | 0.537 | 6 |
| short moving average | -0.317 (0.454) | -0.890 (0.055) | 0.102 | 0.527 | 8 |
| exp. smooth | 0.250 (0.382) | -0.898 (0.054) | 0.130 | 0.545 | 4 |
| arma | -0.075 (0.389) | -0.791 (0.043) | 0.001 | 0.538 | 5 |
| *Wavelet-based measures* | | | | | |
| db10 (level 4) | -0.101 (0.371) | -0.910 (0.055) | 0.004 | 0.547 | 3 |
| sym5 (level 5) | -0.082 (0.354) | -0.910 (0.054) | 0.177 | 0.552 | 2 |
| db2 (level 3) | -0.101 (0.372) | -0.900 (0.055) | 0.184 | 0.512 | 9 |
| db3 (level 5) | -0.088 (0.317) | -0.911 (0.053) | 0.239 | 0.563 | 1 |
| Haar (level 2) | -0.101 (0.388) | -0.808 (0.058) | 0.246 | 0.486 | 10 |
| sym1 (level 4) | -0.090 (0.363) | -0.897 (0.056) | 0.250 | 0.534 | 7 |

*Notes*: First numbers in second and third columns are estimated parameters; numbers in parentheses are their estimated standard errors. Results based on equation (1) with AR(1) errors, estimated using 401 observations over 67:3 to 00:7 for a horizon of 18-months. The F-test is testing the restriction that $\alpha_H = 0$ and $\beta_H = -1$.